# Ripplon Laser

Samuel Kaminski, Leopoldo L. Martin*, Shai Maayani and Tal Carmon.

*Faculty of Mechanical Engineering, Technion -, Haifa, Israel.*

*Corresponding author*

Energy exchange between light and sound was first suggested by Brillouin in 1922[1]. After Townes established the "*phonon-maser*" theory[2], coherent generation of intense hypersonic waves was observed together with stimulated Brillouin Scattering[3,4]. Here we repeat these experiments, but with capillary-waves that are unique to the liquid-phase of matter and relates to attraction between intimate fluid particles. We fabricate resonators that co-host capillary-[5] and optical-modes[6], control them to operates at their non-resolved sideband, and observe stimulated capillary-scattering and coherent excitation of capillary resonances at kHz rates (that one can hear in audio files recorded by us). By exchanging energy between electromagnetic- and capillary-waves, we bridge interfacial-tension phenomena at the liquid-phase boundary to optics, and might impact optofluidics by allowing optical-control, -interrogation and -cooling[7] of water waves.

Novel nanolasers[8-10] were benefiting from going to metallic materials. Similarly, one can go to different wave systems in order to extend the ideas of Brillouin lasers[2-4], relaying on acoustical waves, to capillary waves. At a scale smaller than the width of a hair (Latin: capillus), effects arising from cohesive forces between intimate particles of liquid result in tension that pushes to minimize the liquid surface. Unlike acoustical waves that can propagate in all phases of matter, capillary waves are unique to liquids. Liquid interfaces, such as the ones in our glass of water, tend to be viewed as stationary smooth surfaces that minimally exchange energy with electromagnetic waves. Yet, a liquid-phase boundary, including in droplets, actually looks like a stormy sea when monitored with Angstrom resolution[11,12], and can therefore scatter and Doppler-shift light. During this scattering process, light can transfer part of its energy to capillaries through red Doppler-shift. Similar scattering from density variations while coherently generating acoustic waves is called a Brillouin laser[2-4]. Raman scattering[2,4] refers to scattering from fluctuations in charge distribution while the molecule's center of mass remains stationary. In such scattering processes, energy-conservation considerations suggest stimulated optical generation of coherent oscillations if "redder" (Stokes) scattering governs over the bluer (antiStokes) ones. Thermodynamically and as one can see in figure 1, the Boltzmann's distribution[13] implies a sufficient Stokes enhancement for Raman- and Brillouin-lasers[2]. Quite the opposite, capillaries have rates ($10^3$ Hz) much lower than optical frequencies ($10^{14}$), suggesting that their Stokes and antiStokes lines are almost equal. In other words, there will be no optocapillary energy exchange because the Stokes-created ripplons will be balanced by the antiStokes-annihilated ripplons. To mitigate this Stokes-antiStokes balance, one can "help" the Boltzmann distribution in the task of enhancing the capillary Stokes line. For example, the Stokes line can be resonantly enhanced (over the antiStokes one) by operating at the sideband of an optical resonance (Fig. 1b). Such a usage of sidebands-control is major cooling of gasses[7] and excitation of phonon lasers[14]. It seems therefore controlling a liquid-droplet optical-cavity[6,15-20] to operate at its non-resolved sideband refime will allow ripplon-laser emission. Yet, just like low acoustical attenuation is needed for Brillouin lasers[2-4], low-viscosity is required for capillary oscillations. Low-viscosity (runny) liquids are however experimentally challenging because of their inherently fast evaporation-rate. A recently demonstrated resonator is both small and proof against evaporation as it uses optical tweezers to submerge a microdroplet near an optical coupler[18]. Here we use this technique[18] to cancel evaporation, but use runny liquids that transform the system from the capillary-overdamped to the capillary-underdamped operation regime.

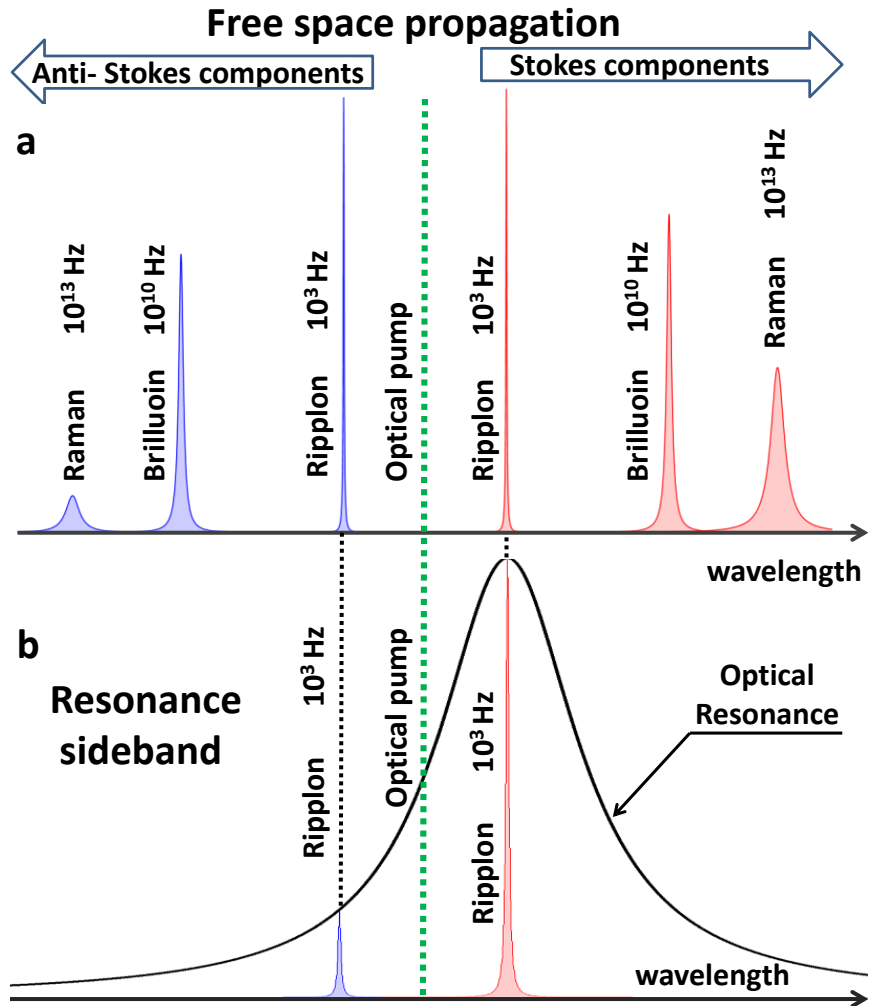

**Figure 1. Light-matter interactions**. (a) In bulks, optical- and acoustical-waves have rates that are large-enough to thermodynamically enhance their Stokes lines to enable Raman and Brillouin lasers. In contrast, the lower frequency of ripplons suggests a Stokes-antiStokes balance that makes ripplon lasers challenging. (b) Mitigating this limitation and tilting this balance to favor Stokes emission is suggested here by working at the non-resolved sideband of an optical cavity.

Our hybrid opticapillary resonator is an octane-microdroplet in water. Octane and water are chosen here for their low viscosity (0.54 and 1.0 mPa·s) and for their high transparency which allows a capillary quality factor ($Q_{cap}$) of 18 together with an optical quality factor ($Q_{opt}$) of 10 million. When the droplets circumference is an integer number of optical wavelengths, an optical whispering-gallery mode (Fig 2a, circular arrow) is possible. The transverse profile of this mode is calculated[21] to resides at the droplet interface (Fig 2a, inset). Similar to its optical modes, the droplet can also capillarlly resonate when its size is near an integer number of capillary-wavelength. The capillary shape-oscillations (Fig. 2b) eigenfrequencies[22], $f_i$, are

(1) $$f_i^2 = \{\gamma(i+1)^2(i+2)^2\} / \{4\pi^2 R^3[(i+2)\cdot\rho_o + (i+1)\cdot\rho_w]\} = k/m \ ,$$

where γ is the surface tension, $R$ is the droplet radius, $\rho_o$ and $\rho_w$ are the mass densities of octane and water, and $i$ is the mode order. To give a scale relevant to our experiment, an R=90 μm droplet with $\rho_w$=1000, $\rho_o$=708 kg/m³ and γ=10.5 mJ/m² will have a natural frequency, $f_1$, of 1784 Hz and $f_3$ of 4400 Hz. As described in[23], k and m here are the effective mass and spring constant for a given droplet mode and are convenient for describing radius changes, r, of the droplet using the harmonic-oscillator equation[23]

(2) $$m\ddot{r}(t) + b\dot{r}(t) + kr(t) = 2\pi|A(t)|^2/c \ ,$$

where b is the damping coefficient, $c$ is the speed of light in the resonator and $|A(t)|^2$ is the optical power circulating in the droplet while inflating its radius by applying centrifugal radiation pressure. Cavity inflation, r(t), will in return Doppler shift the optical resonance frequency, $\Delta\omega(t)$ as described by[23]

(3) $$\Delta\omega(t) = \Delta\omega_0 - \frac{2\pi\omega\cdot r(t)}{\lambda N} \ ,$$

where $\lambda$ is the optical wavelength in the resonator, N is an integer describing the circumference divided by resonance wavelength, and $\omega$ is the angular optical frequency. $\Delta\omega_0$ is the detuning between the frequency of the optical pump and the cavity resonance frequency while at capillary equilibrium. $\Delta\omega_0$ therefore describes the initial sidebanding size and sign. The total detuning, $\Delta\omega(t)$, affects the optical field inside the resonator, $A(t)$ (in $\sqrt{W}$) as described by[24]

(4) $$\dot{A}(t) + A(t)[\alpha c - I\Delta\omega(t)] = IB\sqrt{\frac{\alpha c}{\tau_0}} \ ,$$

where $\alpha$ is the optical loss in the cavity, B is the input pump field, $I=\sqrt{-1}$, and $\tau_0$ is the photon roundtrip time. This set of coupled differential-equations (Eq. 2-4) will later be solved with parameters as in our experiment.

We *fabricate* optocapilary μdroplet resonators by blending octane, water and a "tween-40" emulsifier (14.9, 85, and 0.1%) in an ultrasonic bath. Using a pipette, we then take an octane droplet to our *experimental setup* (Fig 2.a) where optical tweezers position[18] it near a tapered-fiber coupler[25]. Importantly, we fabricate our taper so that the speed of light in it equals to the speed of light circulating in the resonator. This enables phase-matching that can potentially allow for all the light to couple. During the experiment, we can have a finer control over the speed of light by moving the resonator towards thinner (or thicker) regions of the taper.

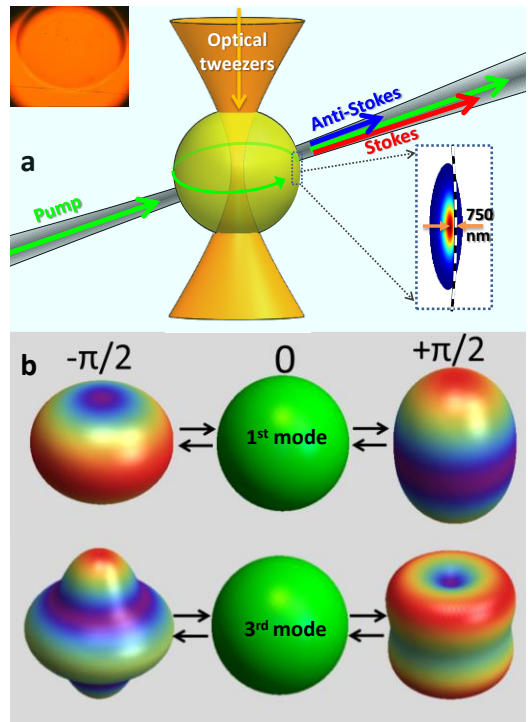

**Figure 2. Experimental setup:** (a) Optical tweezers (orange) manipulate an octane microdroplet (yellow, and inset) submerged in water (light blue) near a tapered coupler from which it receives pump light (green). Radiation pressure by circulating light (green ring) excites capillary oscillations giving rise to Stokes radiation. (b) Droplet capillary resonances[22] (see Supplementary movies 1 and 2).

We will now experimentally demonstrate our major claim of optically exciting capillary resonances. We scan our pump laser through a droplet's optical resonances (Fig 3a). As common to microresonators, the average absorption is shark-tail shaped, originating from thermal effects[26]. Yet unlike in any previous experiment, including those with whispering-gallery resonators, a periodic transmission oscillation is measured at kHz rate. As these recorded oscillations (figure 2a, e) are at audio rates, one can actually listen to them in a soundtrack that we provide (Fig 2b, blue, complementary audio file a-d). Importantly, we finely control the taper-resonator distance (by positioning the optical tweezers) until transmission-oscillations amplitude reaches a maximum. This controlled coupling is significant in extracting the largest portion of the Stokes line out of the resonator, which is commonly referred to as reaching an "optimal coupling"[27]. When optimally coupled (Fig 3a) we observe strong interference oscillations between the Stokes and pump light with an amplitude reaching 70% of the transmission, indicating the strength of the Stokes line. Transmission is calculated with Eq 2-4 and presented next to the experimental data (Fig 3b, red). The parameters in this calculation are $Q_c=18$, as revealed[28] from the known viscosities and mass densities, $Q_{opt}=10^7$, as estimated based on octane transmission, and $\gamma=10.5$ mJ/m$^2$ was extracted from Eq.1.

We now present the transmission-oscillation amplitude vs. the optical power absorbed by the resonator (fig 3d). As common[27], increasing the light-in power from zero is accompanied by a linear increase in the oscillation; indicating spontaneous scatterings from Brownian capillary-fluctuations. A knee at 77.4 µW marks the region where optical gain becomes equal to the energy loss (including by viscosity) which is the power-threshold region for oscillation. Theoretically calculating the power threshold using[29]

$$(5) \quad P_{threshold} = \frac{5 \cdot R^2 \omega^2 \sqrt{m \cdot k}}{32} \frac{1}{Q_{opt}^3 \cdot Q_{cap}}$$

reveals a $Q_{opt}=3 \cdot 10^6$ reveals the measured threshold. At powers larger than threshold, the curve's slope steepens and then saturating; indicating optically induced oscillation and then reaching the depleted-pump regime. As for the oscillations linewidth, it narrows with input power (Fig 3d).

We now repeat this experiment (Fig 3 a-c) while reversing the scanning direction of our laser. By reversing the scan direction we experimentally verify that oscillations appear, again, at the blue sideband (Fig 3e) irrespective of the scan direction.

Eq. 1 suggests that the theoretically expected-ratio[22] between the frequency of the 3$^{rd}$ and the 1$^{st}$ mode is 2.47, which deviates 2.8% from our experimental measurement (4500/1770=2.54, Fig. 3 c, g). Excitation of different modes (Fig 2 c, g) is typical since higher optical-quality modes tend to excite lower-rate oscillations[30]. The absence of the 2$^{nd}$ mode is also typical due the fact that even modes tend to have their node near the equator and therefore have a smaller optocapillary coupling[29].

In conclusion, our optocapillary cavity extends Brillouin-lasers ideas from sound to capillary waves in liquids. Our optical- and capillary-quality-factors were sufficient for demonstrating a ripplon laser. Maybe even more importantly, fine-control over coupling and frequency detuning allows to optimally couple light originating from optocapillary interactions into the fiber while operating at the non-resolved sideband. This level of control and quality-factors might soon extend the impact of optocapillary cavities to establish a new bridge between microfluidics and optics and enabling a playground for nonlinear optocapillary interactions.

Acknowledgements:
This research was supported by ICore: the Israeli Excellence center "Circle of Light" and by the Israeli Science Foundation grant no. 2013/15.

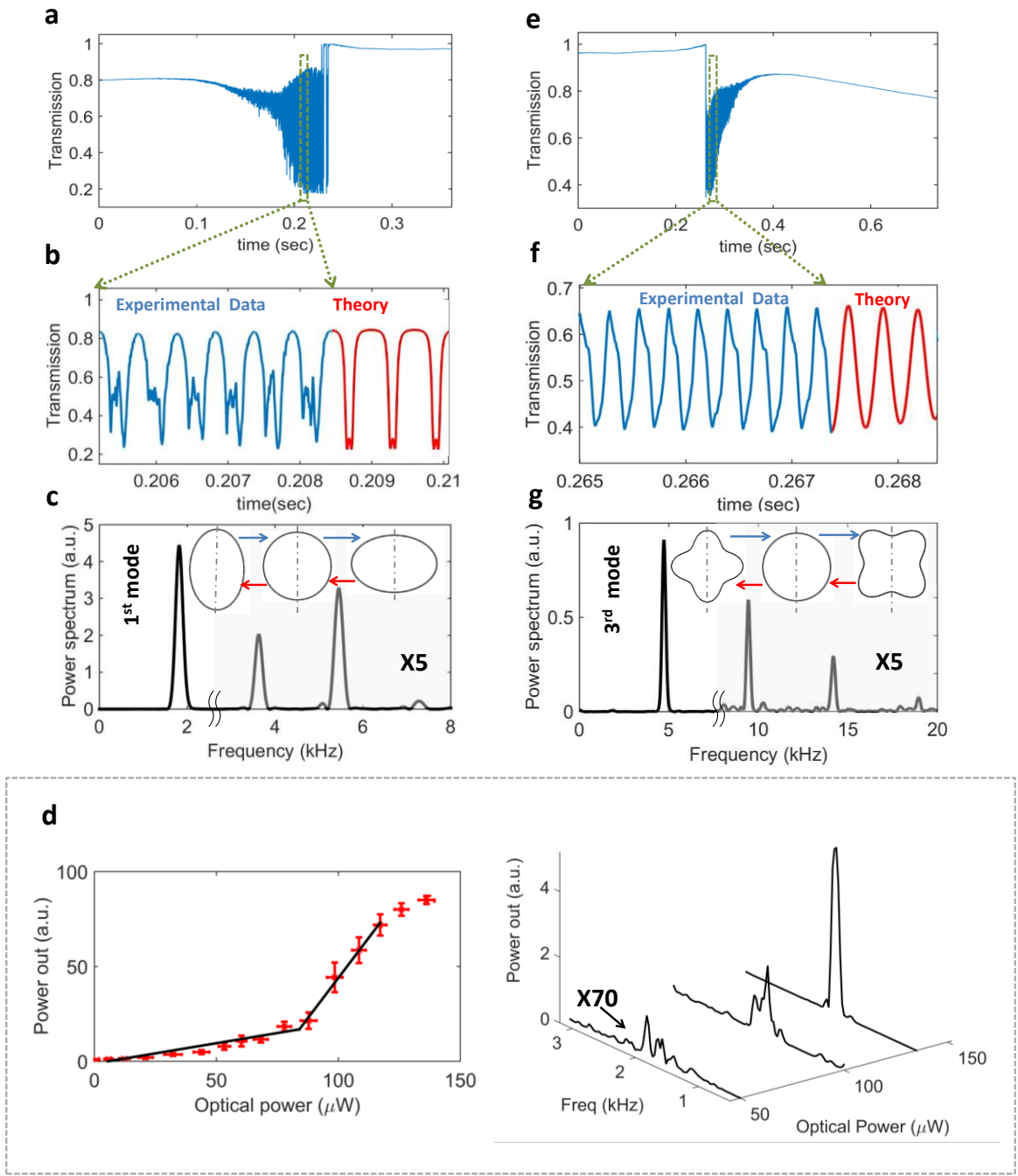

**Figure 3. Experimental results**: Scanning the pump-laser frequency reveals (a-c) periodic oscillations that spectrally narrow (d, right) and grows with input power (d, left) where a knee marks the transition from spontaneous scatterings by Browninan capillary oscillation to optically excited oscillations. (e-g) reversing the scan direction exhibit a mirrored evolution of oscillation (in respect to a) indicating that ripplon laser always appear at the blue sideband irrespective of scan direction. A mode at different frequency is depicted in (e-f). Laser wavelength=770 nm, R=90 μm, refractive index=1.33 for water and 1.4 for octane.

**Supplementary files:**

Supplementary movie 1. Calculation of the droplet $1^{st}$ capillary mode

Supplementary movie 2: Calculation of the droplet $3^{rd}$ capillary mode

Supplementary movies c-f: The movies describe optical transmission while scanning the pump laser through optical resonances. Optical resonances are evident via dips in optical transmission. Some of the resonances are accompanied by transmission oscillations at audio-band frequency that is originating from the optically-induced capillary oscillation. The soundtrack plays these capillary oscillations.

File c: Real time, scanning through ripplon lasers where the Stokes line is 4500 Hz away from the pump.

File d: Real time, scanning through ripplon lasers where the Stokes line is 1770 Hz away from the pump.

File e: 1/10 slow motion, scanning through ripplon lasers where the Stokes line is 4500 Hz away from the pump.

File f: 1/10 slow motion, scanning through ripplon lasers where the Stokes line is 1770 Hz away from the pump.